\documentclass[pra,twocolumn,superscriptaddress]{revtex4}
\usepackage[]{graphicx}
\usepackage{amssymb,amsmath,multirow}

\usepackage{amsfonts,amssymb,amsmath}
\usepackage[]{graphics,graphicx,epsfig}
\usepackage{amsthm}

\begin{document}


\title{The triangle principle: a new approach to non-contextuality and local realism}


\author{Pawe\l\ Kurzy\'nski}
\email{cqtpkk@nus.edu.sg}
\affiliation{Centre for Quantum Technologies,
National University of Singapore, 3 Science Drive 2, 117543 Singapore,
Singapore}
\affiliation{Faculty of Physics, Adam Mickiewicz University,
Umultowska 85, 61-614 Pozna\'{n}, Poland}

\author{Dagomir Kaszlikowski}
\email{phykd@nus.edu.sg}
\affiliation{Centre for Quantum Technologies,
National University of Singapore, 3 Science Drive 2, 117543 Singapore,
Singapore}
\affiliation{Department of Physics,
National University of Singapore, 3 Science Drive 2, 117543 Singapore,
Singapore}


\begin{abstract}
In this paper we study an application of an information distance between two measurements to the problem of non-contextuality and local realism. We postulate {\it the triangle principle} which states that any information distance is well defined on any pair of measurements, even if the two measurements cannot be jointly performed. As a consequence, the triangle inequality for this distance is obeyed for any three measurements. This simple principle is valid in any classical realistic theory, however it does not hold in quantum theory. It allows us to rederive in an astonishingly simple way a large class of non-contextuality and local realistic inequalities via multiple applications of the triangle inequality. We also show that this principle can be applied to derive monogamy relations. The triangle principle is different than the assumption of non-contextuality and local realism, which is defined as a lack of existence of a joined probability distribution giving rise to all measurable data. Therefore, we show that one can design Bell-Kochen-Specker tests using a different principle. 
\end{abstract}


\maketitle


\section{Introduction}

Since the seminal papers by Bell \cite{Bell64} and Kochen-Specker \cite{KS67} we know that quantum mechanics is incompatible with the assumptions of local realism (LR) and non-contextuality (NC). NC/LR hypothesis states that all measurable properties of a physical system do not depend on the context in which they are measured. More precisely, suppose a given physical system has properties $A,B,C$ that yield outcomes $a,b,c$ with some probability distributions $p(a),p(b),p(c)$. Suppose that the property $A$ can be co-measured with the property $B$ giving us a probability distribution $p(a,b)$ or it can be co-measured with the property $C$ giving a probability distribution $p(a,c)$. We say that $A$ can be measured in the context of $B$ or $C$. NC/LR states that there exists a joint probability distribution $p(a,b,c)$ such that $p(a,b)$ and $p(a,c)$ are recovered as marginals. 

Note that it might be impossible to measure $p(a,b,c)$ for some reasons. For instance, in quantum theory, one cannot co-measure two orthogonal components of spin. NC/LR can be extended to more properties $A,B,C,D,\dots$ and is equivalent to the existence of a joint probability distribution (JPD) for all the properties $p(a,b,c,d,\dots)$ \cite{Fine}. 

NC/LR is a very plausible hypothesis based on our everyday experience. The colour of your car, defined by its spectral profile, would be the same regardless if you looked at it together with Prof. Kochen or Prof. Specker. All classical theories of matter are compatible with NC/LR.

In NC/LR tests, context can be established in two different ways. The most common one, which we call Bell scenario, is to assume that $A$ is measured in Alice's laboratory whereas $B$ and $C$, who provide context for $A$, are measured in spatially separated Bob's laboratory. We guarantee the lack of mutual influence between measurements in different laboratories by invoking the fact that information cannot propagate faster than light (in fact, it is enough to assume that there is a finite speed of information propagation). The less common but more general scenario, which we call Kochen-Specker one, is where there is no division into subsystems and all observables are measured on the same system. The context for $A$ is provided by $B,C$ whose lack of influence on $A$ is guaranteed by the so-called no-disturbance assumption. It is therefore clear that LR is a special case of NR and no-signaling is a special case of no-disturbance.

No-disturbance has not been justified by any general principle such as the finite propagation of information but it can be tested for any physical system. Simply measure $A$ alone and then measure $A$ followed by measurement of $B$. Repeat the whole procedure in reverse order. Do the same for $A$ and $C$. If obtained statistics for $A$, i.e., the probability distribution $p(a)$ is the same in all scenarios then you conclude that no-disturbance holds. This way we can verify that quantum mechanics is a no-disturbance theory. Classical theories of matter are, by their very foundations, no-disturbance theories. 
 
It was Bell who showed that LR can be tested experimentally in Bell scenario \cite{Bell64}. Experiments followed \cite{ADR82} clearly demonstrating that quantum mechanics violates LR. The Kochen and Specker \cite{KS67} assertion that indivisible quantum mechanical systems are contextual seemed to be impossible to test experimentally in Kochen-Specker scenario until the paper by Klyachko-Can-Biniciouglu-Schumovsky (KCBS) \cite{KCBS08} whose KCBS inequality was tested experimentally \cite{Lapkiewicz,AACB13}. It is interesting from the sociological point of view that it took 50 years to experimentally test Kochen-Specker theorem whereas Bell scenario was tested within 20 years of its formulation.  

More formally, Bell and Kochen-Specker scenario can be tested via violation of non-contextuality inequalities.  Many such inequalities have been derived for both scenarios. All derivations start from assuming that there is a JPD for all observables used to test a given system. One then manipulates this hypothetical JPD to obtain an expression that involves only measurable marginal probability distributions and it is bounded from below and above by certain bounds resulting from the assumption about the existence of JPD. 

However, inequalities that test LR and NC can be formulated in many ways. For example, the Clauser-Horne-Shimony-Holt (CHSH) inequality \cite{CHSH} uses correlation functions between four measurements, but the same four-measurement scenario can be a base for a derivation of an inequality that utilizes only probabilities \cite{CH}, or give rise to an inequality involving entropies of these measurements \cite{BC,Schumacher,CA}. Each of these inequalities requires a different approach to the JPD problem. It is therefore natural to ask whether there exist some more general approach that treats all types of inequalities on the same footing and that uses some other assumption than the existence of JPD. 

Here, we expand the idea of Schumacher \cite{Schumacher} that Bell inequalities can be expressed as a violation of the properties of the distance measure, and show that such a unification is possible for a large class of inequalities. This can be done under an assumption that an information distance between measurements is well defined on any pair of measurements; Even on measurements that cannot be jointly measured. As a consequence, triangle inequality for a distance measure is always obeyed for any three measurements. Moreover, we show that properties of the distance measure imply a monogamy relation between violations of some of these inequalities. 

As mentioned above, Bell scenario is a special case of Kochen-Specker scenario at least from the mathematical point of view.  There are some researchers who give a special status to Bell scenario claiming that no-disturbance cannot be experimentally verified. They give a simple argument saying that a physical system can have a memory of what context has been used and then return outcomes compatible with no-disturbance hypothesis. It is indeed an open problem but we do not want to discuss it here. Our goal is to study the properties of contextual theories. In this sense, this paper can be considered as a purely mathematical speculation that we nevertheless hope will be useful in deepening our understanding of why Nature is fundamentally non-classical. 


\section{Distance and triangles}

In this section we postulate a new principle that we call {\it the triangle principle}. This principle allows us to derive a large class (we explain later exactly what class it is) of non-contextuality and Bell inequalities as well as their monogamies.


\subsection{Shortest distance form A to B}

\begin{figure}[t]
\begin{center}
\includegraphics[scale=0.24]{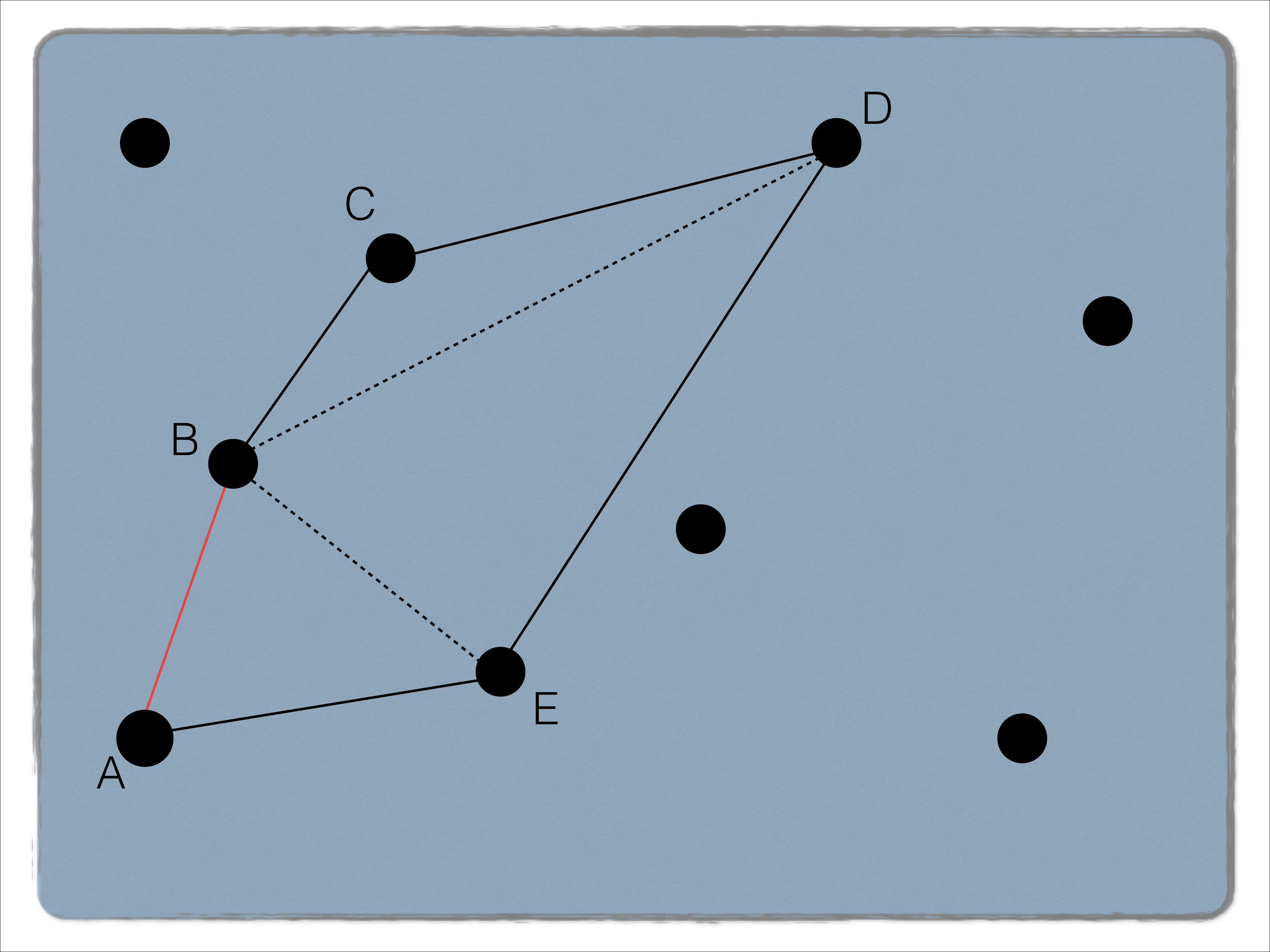}
\end{center}
\caption{\label{f1} In a metric space the shortest distance from $A$ to $B$ is always $d(A,B)$ (red line). This can be shown via multiple application of the triangle inequality. Dashed lines represent distances that do not occur in an alternative path from $A$ to $B$ (solid black lines), but are used to show that the length of this path is never shorter than $d(A,B)$.}
\end{figure}

We start with an abstract metric space with a distance function $d(X,Y)$ between any two points $X$ and $Y$. Consider an arbitrary discrete subset of points $X_1$, $X_2$, ... that belong to this space. It can be easily shown that $d(X_1,X_2)$ is the shortest distance between $X_1$ and $X_2$ and that any other path from $X_1$ to $X_2$ that goes through $N-2$ other points in this subset $X_1 \rightarrow X_N \rightarrow X_{N-1} \rightarrow \dots \rightarrow X_3 \rightarrow X_2$ is never shorter than $d(X_1,X_2)$. More formally,  
\begin{equation}\label{ineq}
d(X_1,X_2) \leq \sum_{i=2}^{N-1}d(X_i,X_{i+1}) + d(X_N,X_1).
\end{equation}
To prove this one simply starts with a triangle inequality
\begin{equation}\label{i2}
d(X_1,X_2) \leq d(X_2,X_N) + d(X_N,X_1).
\end{equation}
Since $d(X_2,X_N)$ does not appear in (\ref{ineq}), one uses another triangle inequality to bound this element from above
\begin{equation}
d(X_2,X_N) \leq d(X_{N-1},X_N) + d(X_2,X_{N-1}),
\end{equation}
which after substitution to (\ref{i2}) gives
\begin{equation}\label{i3}
d(X_1,X_2) \leq d(X_2,X_{N-1}) + d(X_{N-1},X_n) + d(X_N,X_1).
\end{equation}
The above procedure, that utilizes triangle inequality, is applied untill all elements from (\ref{ineq}) are obtained. A schematic picture corresponding to the case of five points is presented in Fig.~\ref{f1}.


\subsection{Information distance}

Let us consider an experiment where we measure some properties of a physical system denoted as $A,B,C,\dots$. Each property $X$ yields an outcome $x$ with probability $p(x)$. We further assume that only certain pairs of properties can be simultaneously measured. For instance, it is possible to obtain the probability distribution $p(a,b)$ for $A$ and $B$ but not for $A$ and $C$ etc. We are not interested in probability distributions involving more than two properties although they might be measurable in the experiment.  

We introduce a distance function $d(X,Y)$ between two probability distributions $p(x)$ and $p(y)$ having a joint probability distribution $p(x,y)$. This function must be 1. non-negative and $d(X,Y)=0$ if and only if $X=Y$ 2. symmetric $d(X,Y)=d(Y,X)$  3. obey the triangle inequality $d(X,Y)+d(Y,Z)\geq d(X,Z)$ for arbitrary probability distributions $p(x,y),p(y,z),p(x,z)$. 

It is instructive to give a few examples of such distances. 

{\it Covariance distance \cite{Schumacher}.} It is defined for binary random variables ($x=\pm 1$) as $C(X,Y)= 1-\sum_{x,y=\pm1}xy~p(x,y) = 1- \langle X Y \rangle.$ 

{\it Entropic distance \cite{Zurek}.} Definition is as follows $E(X,Y)=H(X|Y)+H(Y|X)$ where $H(X|Y)=H(XY)-H(Y)$ is Shannon conditional entropy, therefore $E(X,Y)=2H(XY)-H(X)-H(Y)$. Interestingly, it is also a distance measure if one replaces Shannon entropy by algorithmic entropy \cite{Zurek}.

{\it Kolmogoroff distance.} It reads $K(X,Y)=P(x_0)+P(y_0)-2P(x_0,y_0)$ where $x_0,y_0$ denote some particular events, for instance, for binary variables we could have $x_0=-1,y_0=1$. It is a simple exercise in Venn diagrams to prove that $K$ is a proper distance measure. 


\subsection{The triangle principle}

We are ready now to formulate the triangle principle: 

\vspace{2mm}

{\it A distance measure $d(X,Y)$ is well defined on all marginal distributions $p(X,Y)$ regardless if they can be measured or not.} 

\vspace{2mm}

In the next section we show that this principle unifies correlation, probability and entropic Bell and non-contextuality inequalities for $N$ cyclicaly compatible measurements in a more general framework. 


\section{Applications and unification}

To illustrate this principle we consider non-contextuality inequalities for $N$ binary measurements $X_1,\dots,X_N$ (each $X_i$ takes only two values) that are cyclically compatible, i.e., measurement $X_i$ can be jointly measured with $X_{i+1}$ (modulo $N$) \cite{Ncycles}. First, we explicitly present the cases of $N=3,4,5$ followed by an arbitrary $N$.


\subsection{N=3}

The case $N=3$ is particularly interesting since if $X_1$ is co-measurable with $X_2$, $X_2$ with $X_3$, and $X_3$ with $X_1$, one may think that all three observables are jointly measurable. This is true in quantum mechanics, however one can consider generalized probabilistic theories (GPT) in which pairwise compatible measurements are not jointly compatible \cite{LSW}.

A special version of this problem was studied by Specker, who considered $X_i$ to be three exclusive events \cite{Specker}. We use notation $X_i=1$ to denote that $X_i$ occured, and $X_i=-1$ to denote that it did not occur. Due to exclusivity the following holds:
\begin{equation}\label{Specker}
p(X_1=1)+p(X_2=1)+p(X_3=1) \leq 1.
\end{equation} 
We refer to this inequality as the Specker inequality or the Specker principle. It is obeyed in classical and quantum theory but it can be violated in GPT \cite{LSW,CSW}.

Let us consider three binary measurements $X_i=\pm 1$ and invoke the triangle principle, i.e., assume that there is a well defined information distance between all three measurements. Therefore, the inequality (\ref{ineq}) gives
\begin{equation}\label{n3}
d(X_1,X_2)\leq d(X_2,X_3) + d(X_3,X_1).
\end{equation}
We show that this general inequality is equivalent to a correlation inequality, an entropic inequality, or a probability inequality (Specker's inequality), if one chooses a proper distance function. Although these three inequalities are satisfied in quantum theory, we show that they can be violated in general probabilistic theories (GPT). Therefore, GPT do not obey the triangle principle.


{\it Covariance distance.} The inequality (\ref{n3}) becomes
\begin{equation}
1-\langle X_1 X_2 \rangle \leq 1-\langle X_2 X_3 \rangle + 1-\langle X_3 X_1 \rangle,
\end{equation}
which gives
\begin{equation}\label{n3c}
-\langle X_1 X_2 \rangle + \langle X_2 X_3 \rangle + \langle X_3 X_1 \rangle\leq 1.
\end{equation}
The above inequality is clearly the correlation non-contextuality inequality discussed in \cite{Ncycles}. Although it is obeyed in quantum theory, it can be violated up to 3 in GPT by the following no-disturbance probability distribution
\begin{eqnarray}
p(X_1=+1,X_2=-1)&=&p(X_1=-1,X_2=+1)=1/2, \nonumber \\
p(X_2=+1,X_3=+1)&=&p(X_2=-1,X_3=-1)=1/2, \nonumber \\ 
p(X_1=+1,X_3=+1)&=&p(X_1=-1,X_3=-1)=1/2, \nonumber \\ 
\label{p1}
\end{eqnarray}
with all the remaining probabilities equal to zero. In all further example we only list non-zero probabilities in any given probability distribution.

{\it Entropic distance.} The entropic version of (\ref{n3}) gives
\begin{eqnarray}
& & H(X_1|X_2) + H(X_2|X_1) \leq \label{n3e} \\ 
& & H(X_3|X_2) + H(X_2|X_3) + H(X_1|X_3) + H(X_3|X_1). \nonumber 
\end{eqnarray} 
It is also obeyed in quantum theory, however the GPT no-disturbance distribution 
\begin{eqnarray}
p(X_1=+1,X_2=+1)&=&p(X_1=-1,X_2=-1)=1/4, \nonumber\\
p(X_1=+1,X_2=-1)&=&p(X_1=-1,X_2=+1)=1/4, \nonumber \\
p(X_2=+1,X_3=+1)&=&p(X_2=-1,X_3=-1)=1/2, \nonumber \\ 
p(X_1=+1,X_3=+1)&=&p(X_1=-1,X_3=-1)=1/2. \nonumber \\ 
\label{p2}
\end{eqnarray}
leads to its violation, i.e., one gets a contradiction that $2 \leq 0$. 

Note that the distribution (\ref{p2}) violates the inequality (\ref{n3c}), but the distribution (\ref{p1}) does not violate (\ref{n3e}). However, as was shown by Chaves \cite{Chaves}, distributions that violate correlation inequalities can be mixed with some non-contextual distributions to give distributions that violate entropic inequalities. For example, (\ref{p2}) can be obtain from an even mixture of (\ref{p1}) and a non-contextual distribution, that obeys both (\ref{n3c}) and (\ref{n3e})  
\begin{eqnarray}
p(X_1=+1,X_2=+1)&=&p(X_1=-1,X_2=-1)=1/2, \nonumber\\
p(X_2=+1,X_3=+1)&=&p(X_2=-1,X_3=-1)=1/2, \nonumber \\ 
p(X_1=+1,X_3=+1)&=&p(X_1=-1,X_3=-1)=1/2, \nonumber \\ 
\end{eqnarray}


{\it Kolmogorov distance.} We define three events $A=(X_1=1,X_2=-1)$, $B=(X_2=1,X_3=1)$, $C=(X_3=-1,X_1=-1)$. These events are pairwise exclusive, therefore $p(X=1,Y=1)=0$ (for $X,Y=A,B,C$) and as a consequence $P(X=1,Y=0)=P(X=1)$.

Let us consider the following version of (\ref{n3})
\begin{equation}
K(A=0,B=0) \leq K(B=0,C=1) + K(C=1,A=0).
\end{equation}
Due to exclusivity $K(X=0,Y=0)=p(X=1) + p(Y=1)$, where we used the fact that $p(X=0,Y=0)=1-p(X=1)-p(Y=1)$ and $p(X=0)+p(X=1)=1$. Also, $K(X=1,Y=0)=p(Y=0)-p(X=1)$ due to $p(X=1)=p(X=1,Y=0)$. We get
\begin{equation}
p(A=1)+p(B=1) \leq p(B=0) + p(A=0) -2p(C=1),
\end{equation}
which after substitution  of $p(X=0)=1-p(X=1)$ and division by 2 leads to the Specker's inequality
\begin{equation}
p(A=1)+p(B=1)+p(C=1)\leq 1.
\end{equation}
This inequality is violated up to $3/2$ by the GPT distribution (\ref{p1})


\subsection{N=4}

This case naturally describes a bipartite Bell scenario in which Alice measures $X_1$ and $X_3$ while Bob measures $X_2$ and $X_4$. Moreover, $N=4$ is the minimal number of measurements leading to violation of any non-contextual inequalities in quantum theory. The corresponding inequality (\ref{ineq}) reads
\begin{equation}\label{n4}
d(X_1,X_2)\leq d(X_1,X_4) + d(X_4,X_3) + d(X_3,X_2). 
\end{equation}

The distances $d(X_1,X_3)$ and $d(X_2,X_4)$ cannot be evaluated due to the lack of co-measurability. However we use the triangle principle which assumes that these distances, although unaccessible, are well defined. 

We show that depending on the distance function the inequality (\ref{n4}) becomes the Clauser-Horne-Shimony-Holt (CHSH) inequality \cite{CHSH}, the Schumacher inequality \cite{Schumacher}, or the Clauser-Horne (CH) inequality \cite{CH}. All three inequalities can be violated in qunatum theory provided a quantum state and measurement setups are properly chosen. 


{\it Covariance distance.} For covariance distance the inequality (\ref{n4}) becomes 
\begin{equation}
1-\langle X_1 X_2\rangle \leq 3 - \langle X_1 X_4\rangle - \langle X_4 X_3\rangle - \langle X_3 X_2\rangle, 
\end{equation}
which has the form of the CHSH inequality
\begin{equation}\label{n4c}
\langle X_1 X_4\rangle + \langle X_4 X_3\rangle + \langle X_3 X_2\rangle -\langle X_1 X_2\rangle \leq 2. 
\end{equation}
This observation was already made by Schumacher \cite{Schumacher}.


{\it Entropic distance.} Application of the entropic distance to (\ref{n4}) results in the Schumacher inequality, which was already defined using a distance approach
\begin{equation}
E(X_1,X_2)\leq E(X_1,X_4) + E(X_4,X_3) + E(X_3,X_2). 
\end{equation}
It is also important to mention similar entropic inequalies due to Braunstein-Caves (BC) \cite{BC} and due to Cerf-Adami (CA) \cite{CA}. The BC inequality uses conditional entropies, whereas the CA inequality uses mutual information. 


{\it Kolmogorov distance.} We define four events $X_1=1$, $X_2=1$, $X_3=1$ and $X_4=1$. The Kolmogorov distance and the corresponding inequality (\ref{n4}) yields
\begin{eqnarray}
K(X_1=1,X_2=1) &\leq& K(X_2=1,X_3=1)+ \nonumber \\ 
K(X_3=1,X_4=1) &+& K(X_4=1,X_1=1),
\end{eqnarray}
which, after substitution $K(X=1,Y=1)=p(X=1)+p(Y=1)-2P(X=1,Y=1)$ and division by 2, takes the form of the CH inequality \cite{CH}
\begin{eqnarray}
&-& p(X_1=1,X_2=1) + p(X_2=1,X_3=1) \nonumber \\
&+& p(X_3=1,X_4=1) + p(X_4=1,X_1=1) \nonumber \\
&-& p(X_3=1) - p(X_4=1)\leq 0.
\end{eqnarray}


\subsection{N=5}

The case of five cyclicaly compatible measurements is often related to the non-contextuality tests because $N=5$ is the smalest number of measurements that can reveal contextuality in a three-level quantum system. Moreover, these measurements cannot be naturaly distributed between two observers, therefore they are applied to a single indivisible system. 

This time there are ten possible distances, but only five of them are measurable and are involved in the inequality (\ref{ineq})
\begin{eqnarray}
d(X_1,X_2)&\leq& d(X_1,X_5) + d(X_5,X_4) \nonumber \\ 
&+& d(X_4,X_3) + d(X_3,X_2). \label{n5}
\end{eqnarray}
The remaining five distances cannot be measured, however the triangle principle assumes that they are properly defined and that the inequality (\ref{n5}) holds. We show that this inequality gives rise to correlation and probability versions of the Klyachko-Can-Binicioglu-Shumovsky (KCBS) inequalities \cite{KCBS} and to the entropic inequality that is similar to the one studied in Refs. \cite{Us,ChF}.


{\it Covariance distance.} Plugging in the covariance distance into (\ref{n5}) results in
\begin{eqnarray}
1- \langle X_1 X_2 \rangle &\leq& 4 - \langle X_1 X_5 \rangle - \langle X_5 X_4 \rangle \nonumber \\ 
&-& \langle X_4 X_3 \rangle - \langle X_3 X_2 \rangle,
\end{eqnarray}
which is equivalent to a form of the KCBS inequality \cite{Ncycles,KCBS}
\begin{equation}\label{n5c}
\langle X_1 X_5 \rangle + \langle X_5 X_4 \rangle +  \langle X_1 X_5 \rangle + \langle X_5 X_4 \rangle - \langle X_1 X_2 \rangle \leq 3. 
\end{equation}


{\it Entropic distance.} As in the case $N=4$, for the entropic distance one gets an entropic inequality that is a five measurement version of the Schumacher inequality. This inequality resembles the inequality studied in \cite{Us,ChF} where instead of $E(X,Y)$ one uses conditional entropy $H(X|Y)$.


{\it Kolmogorov distance.} Application of the Kolmogorov distance to the case $N=5$ resembles the one of $N=3$. Consider five pairwise exclusive events $A=(X_1=1,X_2=-1)$, $B=(X_2=1,X_3=-1)$, $C=(X_3=1,X_4=-1)$, $D=(X_4=1,X_5=-1)$, $E=(X_5=1,X_1=-1)$. As before, $p(X=1,Y=1)=0$ (for $X,Y=A,\dots,E$) and as a consequence $P(X=1,Y=0)=P(X=1)$.

We apply the Kolmogorov distance to (\ref{n5}) and follow exactly the same steps as for $N=3$. We arrive at 
\begin{eqnarray}
& &P(A=1) + P(B=1) + P(C=1) \nonumber \\ &+& P(D=1) + P(E=1) \leq 2,
\end{eqnarray}
which is a version of the KCBS inequality expresed in terms of probabilities \cite{KCBS}.


\subsection{General N}

The discussion of cases $N=3,4,5$ shows that application of the covariance and the entropic distances readily generates correlation and entropic inequalities for general $N$. The correlation inequalities that are generated are of the form
\begin{equation}\label{gnc}
-\langle X_1 X_2 \rangle + \sum_{i=2}^{N} \langle X_{i} X_{i+1} \rangle \leq N-2,
\end{equation}
where $X_{N+1}\equiv X_1$. They exactly correspond to inequalities discussed in \cite{Ncycles}. The entropic inequalities
\begin{equation}\label{gne}
E(X_1,X_2) \leq \sum_{i=2}^{N} E(X_{i},X_{i+1}),
\end{equation}
are $N$ element versions of Schumacher inequalities \cite{Schumacher} and resemble N-cycle conditional entropic inequalities studied in \cite{ChF}. In fact, these inequalities are symmetrized versions of conditional entropic inequalities.

Note that our model also applies to a bipartite Bell scenario if $N$ is even and measurements $X_{2i+1}$ are performed by Alice whereas $X_{2i}$ are performed by Bob. In this case there is an additional number of distinces that can be evaluated from the experimental data. Namely, every distance between Alice's and Bob's measurement is well defined. In this Bell scenario the inequalities (\ref{gnc}) and (\ref{gne}) correspond to the chained Bell inequalities \cite{CBC} and to the symmetric version of the multi-setting BC inequalities \cite{BC}, respectively.    

On the other hand, the form of probability inequalities that are generated via application of the Kolmogorov distance depend on whether $N$ is even or odd. For odd $N$ one can generate inequalities that involve $N$ cyclicaly exclusive events $A_1,\dots,A_N$ defined as $A_i=(X_i=+1,X_{i+1}=-1)$. Next, one considers the Kolmogorov distance for $A_1=0$, $A_i=1$ for $i=3,5,\dots, N$ and $A_i=0$ for $i=2,4,\dots,N-1$. The following inequalities are obtained  
\begin{equation}
\sum_{i=1}^N p(A_1=1) \leq \frac{N-1}{2}.
\end{equation}
They correspond to the inequalities studied in \cite{CSW}.

The case of even $N$ has to be explored in more details. The inequalities involving $N$ cyclicaly exclusive events can be violated in quantum theory and in GPT only for odd $N$ \cite{CSW}. We showed that for $N=4$ one can obtain a different type of inequality, namely the CH inequality. However, the application of Kolmogorov distance to scenarios with $N>4$ (even) remains to be explored.
 

\section{Monogamy relation}

The monogamy relation between two non-contextuality (or Bell) inequalities can be explained as a trade-off between the violations of these inequalities. The more the first inequality is violated the less is the second. In the most interesting case, if one inequality is violated the other is satisfied and vice versa.

The monogamy relation can be studied either on a general level of GPT \cite{Monogamy1,Monogamy2,KCK}, or within quantum theory \cite{KCK,TV06,Monogamy3}. Monogamies in GPT are resulting from general principles like no-signaling and no-disturbance that refer to probabilities, whereas in quantum theory they derive from the very properties of operators in the Hilbert space.

Here we focus ont the GPT case. In particular, we show that instead of refering directly to probabilities (no-signaling/no-disturbance), one can refer to general properties of a distance (triangle inequality) to derive monogamy relations. The advantage of this approach is that once a monogamy relation is derived for a general distance, it authomaticaly applies to every distance measure. We show how to derive a monogamy relation between a Bell inequality ($N=4$) and non-contextuality inequality ($N=5$) \cite{KCK} and a monogamy relation between two bipartite Bell inequalities ($N=4$) \cite{TV06}, however we speculate that our method can be easily applied to more general cases.

 
\subsection{Monogamy between nonlocality and contextuality}

It was shown in \cite{KCK} that there exists a monogamy trade-off between KCBS and CHSH inequalities (corresponding to the covariance distance). Here we show that this result can be generalized to an arbitrary distance function $d(X,Y)$.

Consider two parties Alice and Bob sharing a bipartite system. Alice has five cyclicaly compatible measurements on her subsystem $\{A_1,\dots,A_5\}$ and she randomly choses to perform two of them $A_i$ and $A_{i+1}$ (modulo 5). Bob can perform one of two measurements $B_1$ or $B_2$ on his subsystem. 

Alice's measurements can be used to test the following distance inequality (non-contextuality inequality if one assumes JPD)
\begin{equation}\label{iq1}
d(A_1,A_5) \leq d(A_1,A_2)+d(A_2,A_3)+d(A_3,A_4)+d(A_4,A_5).
\end{equation} 
On the other hand, Bob's measurements and two incompatible measurements of Alice (say $A_1$ and $A_3$) can be used to test another distance inequality (Bell inequality if one assumes JPD)
\begin{equation}\label{iq2}
d(A_1,B_1) \leq d(A_1,B_2)+d(B_2,A_3)+d(A_3,B_1).
\end{equation}

Next, we show that if one inequality is violated, the other one is necessarily obeyed. In particular, it is enough to use the fact that the triangle inequality is always obeyed for compatible measurements to show that the following must hold
\begin{eqnarray}
& & d(A_1,A_5) + d(A_1,B_1) \leq d(A_1,A_2) \nonumber \\ &+& d(A_2,A_3) + d(A_3,A_4) + d(A_4,A_5) \nonumber \\ &+& d(A_1,B_2) + d(B_2,A_3)+d(A_3,B_1). \label{mono}
\end{eqnarray} 
We start with a Triangle inequality
\begin{equation}
d(A_1,A_5) \leq d(A_1,B_2) + d(A_5,B_2),
\end{equation}
and then we expand the last term using another triangle inequality
\begin{equation}
d(A_1,A_5) \leq d(A_1,B_2) + d(A_4,A_5) + d(A_4,B_2).
\end{equation}
We repeat this procedure one more time to obtain
\begin{equation}\label{m1}
d(A_1,A_5) \leq d(A_1,B_2) + d(A_4,A_5) + d(A_3,A_4) + d(A_3,B_2).
\end{equation}
Next, we follow similar steps to obtain
\begin{eqnarray}
d(A_1,B_1) &\leq& d(A_1,A_2)+d(A_2,B_1) \label{m2} \\
&\leq& d(A_1,A_2)+d(A_2,A_3)+d(A_3,B_1). \nonumber
\end{eqnarray}
Finaly, we sum (\ref{m1}) and (\ref{m2}) to get (\ref{mono}).

 
\subsection{Monogamy between two Bell inequalities}

Consider three parties (Alice, Bob and Charlie) sharing a tripartite system. Each of them performs one of two measurements on their subsystems $A_{1,2}$, $B_{1,2}$ and $C_{1,2}$. Note, that due to spacelike separation the measurements $A_i$, $B_j$ and $C_k$ ($i,j,k=1,2$) are mutualy compatible. 

Next, consider two Bell inequalities
\begin{eqnarray}
d(A_1,B_1) \leq d(A_1,B_2)+d(B_2,A_2)+d(A_2,B_1), \label{bm1} \\
d(A_1,C_1) \leq d(A_1,C_2)+d(C_2,A_2)+d(A_2,C_1). \label{bm2}
\end{eqnarray}
Using the same methods as in the previous example we can show that 
\begin{equation}
d(A_1,B_1) \leq d(A_1,C_2)+d(C_2,A_2)+d(A_2,B_1) 
\end{equation}
and
\begin{equation}
d(A_1,C_1) \leq d(A_1,B_2)+d(B_2,A_2)+d(A_2,C_1). 
\end{equation}
The sum of these two inequalities gives the monogamy relation
\begin{eqnarray}
d(A_1,B_1) + d(A_1,C_1) \leq d(A_1,B_2) + d(B_2,A_2) \nonumber \\ + d(A_2,B_1) + d(A_1,C_2) + d(C_2,A_2) + d(A_2,C_1). \label{mono2}
\end{eqnarray} 


\section{Conclusions}

The triangle principle is an assumption about Nature on the same footing as the NC/LR assumption. Both assume some mathematical properties of observed and un-observed probability distributions in Nature. NC/LR assumes an existence of a hypothetical JPD that cannot be obtained in an experiment. The triangle principle assumes an existence of a function assigning numbers corresponding to distances between non-measurable properties.

Mathematicaly speaking, NC/LR treats measurements and outcomes as points in a space with a measure (probability), whereas the triangle principle introduces a metric to this space which allows us to study the relation between these points using a geometric intuition. As we showed, this allows us to unify different types of non-contextuality and Bell inequalities in a more general framework and to derive more general monogamy relations.

We would like to highlight once more that {\it the triangle principle} leads to distance inequalities that are mathematically identical to known Bell and non-contextual inequalities but they are based on different assumptions. Therefore, their violations do not imply contextuality in Kochen-Specker scenario and lack of local realism in Bell scenario. Simply speaking they imply that the assumption about the validity of the triangle inequality for non-observed probability distributions is false.    

{\it Acknowledgements.} This work is supported by the Foundational Quesion Institute (FQXi) and by the National Research Foundation and Ministry of Education in Singapore. The authors acknowledge discussions with Marek \.Zukowski, Antoni W\'ojcik, Andrzej Grudka, Jayne Thompson, Luigi Vacanti. 




\begin{thebibliography}{99}


\bibitem{Bell64}
 J. S. Bell,
 Physics \textbf{1}, 195 (1964).

\bibitem{KS67}
 S. Kochen and E. P. Specker,
 J. Math. Mech. \textbf{17}, 59 (1967).

\bibitem{Fine}
 A. Fine,
 Phys. Rev. Lett. {\bf 48}, 291 (1982).

\bibitem{ADR82}
 A. Aspect, J. Dalibard, and G. Roger,
 Phys. Rev. Lett. \textbf{49}, 1804 (1982).

\bibitem{KCBS08}
 A. A. Klyachko, M. A. Can, S. Binicio\u{g}lu, and A. S. Shumovsky,
 Phys. Rev. Lett. {\bf 101}, 020403 (2008).

\bibitem{Lapkiewicz}
 R. {\L}apkiewicz, P. Li, C. Schaeff, N. Langford, S.~Ramelow, M.~Wie\'sniak, and A.~Zeilinger,
 Nature (London) \textbf{474}, 490 (2011).

\bibitem{AACB13}
 J. Ahrens,
 E. Amselem,
 A. Cabello, and
 M. Bourennane,
 Sci. Rep. \textbf{3}, 2170 (2013).

\bibitem{CHSH}
J. F. Clauser, M. A. Horne, A. Shimony, and R. A. Holt, 
Phys. Rev. Lett. \textbf{23}, 880 (1969).

\bibitem{CH}
J. F. Clauser and M. A. Horne,
Phys. Rev. D \textbf{10}, 526 (1974).


\bibitem{BC}
S. L. Braunstein and C. M. Caves,
Phys. Rev. Lett. \textbf{61}, 662 (1988).

\bibitem{Schumacher}
B. Schumacher,
Phys. Rev. A \textbf{44}, 7047 (1991).

\bibitem{CA}
N. J. Cerf and C. Adami,
Phys. Rev. A \textbf{55}, 3371 (1997).

\bibitem{Zurek}
W. H. Zurek,
Nature \textbf{341}, 119 (1989).

\bibitem{Ncycles}
M. Araujo, M. T\'ulio Quintino, C. Budroni, M. Terra Cuhna, and A. Cabello, 
Phys. Rev. A \textbf{88}, 022118 (2013).

\bibitem{LSW}
Y.-C. Liang, R. W. Spekkens, and H. M. Wiseman,
Phys. Rep. {\bf 506}, 1 (2011).

\bibitem{Specker}
E.P. Specker,
Dialectica {\bf 14}, 239 (1960).

\bibitem{CSW}
A. Cabello, S. Severini, and A. Winter,
arXiv:1010.2163 (2010).

\bibitem{Chaves}
R. Chaves,
Phys. Rev. A {\bf 87}, 022102 (2013).

\bibitem{KCBS}
A. A. Klyachko, M. A. Can, S. Binicioglu, and A. S. Shumovsky, 
Phys. Rev. Lett. {\bf 101}, 020403 (2008).

\bibitem{Us}
P. Kurzy\'nski, R. Ramanathan, and D. Kaszlikowski,
Phys. Rev. Lett. \textbf{109}, 020404 (2012).

\bibitem{ChF}
R. Chaves and T. Fritz,
Phys. Rev. A \textbf{85}, 032113 (2012).

\bibitem{CBC}
S. L. Braunstein and C. M. Caves, 
Ann. Phys. (N.Y.) {\bf 202}, 22 (1990).

\bibitem{Monogamy2}
 R. Ramanathan, A. Soeda, P. Kurzy\'nski, and D. Kaszlikowski,
 Phys. Rev. Lett. \textbf{109}, 050404 (2012).

\bibitem{Monogamy1}
 M. Pawlowski and \v{C}. Brukner,
 Phys. Rev. Lett. \textbf{102}, 030403 (2009).

\bibitem{Monogamy2}
R. Ramanathan, A. Soeda, P. Kurzynski, and D. Kaszlikowski,
Phys. Rev. Lett. {\bf 109}, 050404 (2012).

\bibitem{KCK}
P. Kurzynski, A. Cabello, D. Kaszlikowski,
arXiv:1307.6710 (2013).

\bibitem{TV06}
 B. Toner and F. Verstraete,
 \eprint{arXiv:quant-ph/0611001}.

\bibitem{Monogamy3}
 P. Kurzy\'nski, T. Paterek, R. Ramanathan, W. Laskowski, and D. Kaszlikowski,
 Phys. Rev. Lett. \textbf{106}, 180402 (2011).




\end{thebibliography}
\end{document}